\documentclass[conference]{IEEEtran}
\IEEEoverridecommandlockouts
\usepackage{cite}
\usepackage{amsmath,amssymb,amsfonts,bm}
\usepackage{algorithmic}
\usepackage{graphicx}
\usepackage{textcomp}
\usepackage{xcolor}
\usepackage{microtype}                 % better line breaking overall
\usepackage[hyphens]{url}              % allow breaks at hyphens
\usepackage{xurl}                      % allow breaks almost anywhere in URLs
\usepackage[colorlinks=true,allcolors=blue,breaklinks=true]{hyperref}
\usepackage{cleveref}
\providecommand{\doi}[1]{}% suppress printing of DOIs
\def\BibTeX{{\rm B\kern-.05em{\sc i\kern-.025em b}\kern-.08em
    T\kern-.1667em\lower.7ex\hbox{E}\kern-.125em X}}

\newcommand{\cM}{\mathcal{M}}

\newcommand{\cR}{\mathcal{R}}
\newcommand{\cS}{\mathcal{S}}

\renewcommand{\Pr}{\mathbb P}

\def\eps{{\varepsilon}}
\renewcommand{\vec}{\bm}

\renewcommand{\Pr}{\mathbb P}

\begin{document}

\title{Privacy at Scale in Networked Healthcare}

\author{
\IEEEauthorblockN{M. Amin Rahimian}
\IEEEauthorblockA{\textit{Industrial Engineering} \\
\textit{University of Pittsburgh}\\
Pittsburgh, PA, USA \\
rahimian@pitt.edu}
\and
\IEEEauthorblockN{Benjamin Panny}
\IEEEauthorblockA{\textit{Intelligent Systems Program} \\
\textit{University of Pittsburgh}\\
Pittsburgh, PA, USA \\
bmp83@pitt.edu}
\and
\IEEEauthorblockN{James B.D. Joshi}
\IEEEauthorblockA{\textit{Informatics \& Networked Systems} \\
\textit{University of Pittsburgh}\\
Pittsburgh, PA, USA \\
jjoshi@pitt.edu}
}

\maketitle

\begin{abstract}
Digitized, networked healthcare promises earlier detection, precision therapeutics, and continuous care; yet, it also expands the surface for privacy loss and compliance risk. We argue for a shift from siloed, application-specific protections to \emph{privacy-by-design at scale}, centered on decision-theoretic differential privacy (DP) across the full healthcare data lifecycle; network-aware privacy accounting for interdependence in people, sensors, and organizations; and \emph{compliance-as-code} tooling that lets health systems share evidence while demonstrating regulatory due care. We synthesize the privacy-enhancing technology (PET) landscape in health (federated analytics, DP, cryptographic computation), identify practice gaps, and outline a deployable agenda involving privacy-budget ledgers, a control plane to coordinate PET components across sites, shared testbeds, and PET literacy, to make lawful, trustworthy sharing the default. We illustrate with use cases (multi-site trials, genomics, disease surveillance, mHealth) and highlight \emph{distributed inference} as a workhorse for multi-institution learning under explicit privacy budgets.
\end{abstract}


\section{Future of Privacy in Networked Healthcare}

\bstctlcite{IEEEexample:BSTcontrol}

 The essential function of healthcare thrives on the provider-patient trust, which would be undermined if patient privacy is violated \cite{nass2009value}. Maintaining strong privacy guarantees also aids in the protection of patient data against misuse for profit, ransoming, discriminatory practices, and more. Data will take a central role in future medical AI powered by the explosive growth of digitized healthcare data and data-driven computational frameworks that can improve quality and reduce healthcare costs. However, this data-centric future brings about new challenges in ensuring patient privacy and calls for a foundational privacy framework that can be uniformly applied to different components of the healthcare system.

The growing availability of health data, together with advances in machine learning and artificial intelligence, opens significant opportunities to reshape health sciences and medicine \cite{rajpurkar2022ai}. Data-driven computational frameworks have permeated all fields of health sciences and medicine, from digital contact tracing \cite{ferretti2020quantifying,ferretti2024digital} and the promotion of mental health and psychological well-being \cite{li2023systematic} to medical image analysis \cite{yang2024limits}, cancer diagnosis \cite{chanda2024dermatologist,esteva2017dermatologist}, identifying diabetic retinopathy \cite{gargeya2017automated}, arrhythmia detection \cite{hannun2019cardiologist}, surgery \cite{chadebecq2023artificial}, and trauma care \cite{cannon2024digital}. Major healthcare trends such as smart and digital health \cite{prahalad2024equitable,chen2023digital} and precision medicine \cite{hodson2016precision}, as well as nascent ideas such as AI-driven behavior change, promise to improve health outcomes and enable individualized and democratized healthcare delivery using digital technologies that build on high-resolution biometric data (e.g., heart rate, respiration rate, blood pressure, and oxygen levels) and biomedical data (e.g., genomics and proteomics) as well as lifestyle and environmental measurements (e.g., through wearable sensors), combined with AI-enabled analytics for detection, classification, diagnosis, and care optimization. Realizing these positive social outcomes is contingent on the trust of patients in sharing their digital health data with healthcare providers \cite{shandhi2024assessment,miller2022privacy}.

Data concerning health are considered sensitive (e.g., under GDPR article 9 \cite{gdpr_text_2016}), requiring special protection, as their exposure may violate essential rights and individual freedom and subject people to various forms of discrimination in employment, insurance and services, as well as abuse or attacks. Information sharing between patients and providers is essential to the healthcare care function and is protected by professional secrecy (rooted in the Hippocratic Oath). However, 
 With the increasing complexity of the healthcare system and the explosive growth and availability of digital healthcare data that permeate all aspects of healthcare operations, there is a need for holistic methods that can mitigate privacy risks and balance the societal benefits of public health and improved medical care against the risks to individuals.  

Regulations such as the Health Insurance Portability and Accountability Act (HIPAA), Genetic Information Nondiscrimination Act (GINA), and Federal Policy for the Protection of Human Subjects (Common Rule) have emerged to moderate data governance and mitigate privacy issues related to health data. However, the rapidly changing landscape of digital health and the contextual nature of informational norms in the healthcare domain require flexible, expressive, and adaptable frameworks to evaluate the privacy-utility trade-offs that emerging technologies enable \cite{nissenbaum2004privacy}, in a holistic way throughout the healthcare cycle and across the data collection and algorithmic pipelines that enable public health protocols and healthcare delivery. In laying out a vision for scalable healthcare privacy, we are motivated, on the one hand, by the need to share data broadly to maximize their use for the advancement of medicine and health sciences and the increasing adoption of data-driven interventions for the screening, diagnosis, prognosis, treatment, and monitoring throughout the healthcare cycle and in public health protocols; and, on the other hand, by the need to balance the societal benefits of monitoring and controlling disease, improving healthcare and advancing health sciences against the individual rights to privacy, dignity, and self-determination, as well as protection from undue harms of bias and discrimination on the basis of shared health data.

\subsection{Aligning with National Priorities and Strategic Initiatives}
The ability to privately share and analyze data in healthcare settings and beyond aligns with the “Executive Order on the Safe, Secure, and Trustworthy Development and Use of Artificial Intelligence,” which emphasizes the role of privacy-enhancing technologies (PETs) in a responsible AI future \cite{ai-eo}, and with the earlier \textit{National Strategy to Advance Privacy-Preserving Data Sharing and Analytics}, which calls for a coordinated portfolio of PETs, pilots, standards, testbeds, usability, workforce development, and international cooperation  to ``\textit{advance the well-being and prosperity of individuals and society, and promote science and innovation in a manner that affirms democratic values}'' \cite{national-strategy-ppdsa}. 

Complementary federal roadmaps sharpen this agenda for the research community: the \textit{National Privacy Research Strategy} prioritizes multidisciplinary work to reduce privacy risks from data analytics and AI and to advance machine-readable policy and accountability \cite{nprs-2025}; the \textit{Federal Cybersecurity R\&D Strategic Plan} highlights trust models that integrate identity, access, interoperation, and measurable privacy risk minimization \cite{federal-cyber-rd-2023}; the \textit{NIST AI Risk Management Framework} and the \textit{Blueprint for an AI Bill of Rights} foreground human-centered transparency, safety, and fairness considerations for AI systems that process sensitive data \cite{nistairmf23,ai-bill-of-rights-2022}; and the \textit{NIST Privacy Framework} provides an enterprise risk-management lens to evaluate privacy, security, and utility trade-offs across the data life-cycle \cite{nistpf20}. These strategies suggest a community road-map: end-to-end evaluation and benchmarking of PETs, policy-aware data and model life-cycles, usability and transparency by design, rigorous trade-off metrics, and shared testbeds that connect research to deployment and emerging standards \cite{national-strategy-ppdsa,nprs-2025,federal-cyber-rd-2023}.

\subsection{Adapting to an Evolving Regulatory \& Policy Landscape}
Rules for health data use are shifting in ways that both enable and constrain learning health systems. In research, NIH’s Data Management \& Sharing policy (effective Jan.\ 25, 2023) expands expectations for reuse, making privacy-by-design methods and auditable lifecycles first-order needs rather than add-ons \cite{NIH_DMS_2021,NIH_DMS_Reminder_2023}. Beyond HIPAA, the FTC’s updated \emph{Health Breach Notification Rule} brings many consumer health apps and connected devices under clearer breach-notice duties, tightening accountability across the digital-health ecosystems that interoperate with clinical workflows and community health \cite{FTC_HBNR_PR_2024,FTC_HBNR_FR_2024,FTC_HBNR_Basics_2024}. On the clinical IT side, Office of the National Coordinator for Health Information Technology (ONC)’s recent rule (HTI-1) adds transparency and safety obligations for predictive decision-support in certified health IT, making auditable model lifecycles a practical requirement  and strengthening information sharing \cite{ONC_HTI1_FR_2024,HTI1_Explainers_2024}. 

At network scale, The Trusted Exchange Framework and Common Agreement (TEFCA) is a national initiative in the United States that facilitates secure, seamless, and nationwide exchange of electronic health information (EHI) across disparate health information networks (HINs), mandated by the 21st Century Cures Act. TEFCA’s designated and candidate QHINs (quialified HINs)  are maturing nationwide exchange pathways, expanding the surface where compliance-as-code (privacy budget ledgers, provenance, machine-readable policies) can provide due care while enabling cross-institution learning \cite{TEFCA_QHINs}. At the same time, litigation in 2025 vacating HHS’s 2024 reproductive-health privacy amendments to HIPAA underscores cross-jurisdictional uncertainty around permissible Protected Health Information (PHI) flows and the importance of transportable privacy guarantees \cite{HIPAA_Repro_FactSheet_2025,HIPAA_Repro_Vacatur_2025,Reuters_Vacatur_2025}. 

Internationally, the EU AI Act (OJ 12 July 2024) classifies many health-care AI uses as “high-risk,” imposing lifecycle risk-management and transparency duties that will shape trans-Atlantic collaborations, while GDPR’s constraints on secondary use and cross-border transfers continue to affect clinical research \cite{EU_AI_Act_OJ_2024,MRCT_GDPR_2024}. Concurrently, emerging biometric rules (e.g., strengthened notice and consent requirements, retention limits, and private enforcement in several states) reinforce this direction by rewarding designs that minimize raw signal retention and keep sensitive features at the edge: California treats biometrics as “sensitive personal information” under the CCPA/CPRA, Washington includes a biometric-identifiers law with the My Health My Data Act, and Illinois amended their Biometric Information Privacy Act (BIPA) in 2024 while preserving private enforcement, and these trends are reinforced by record Texas settlements enforcing biometric laws against major platforms \cite{CA_CPRA_SPI_1798_140,WA_Biometric_Identifiers,WA_MHMDA,BIPA_SB2979_2024,TX_Meta_1p4B_2024,TX_Google_1p4B_2025}. At the federal level, the FTC’s biometric policy statement articulates unfair use theories for model training and deployment (based on failure to assess foreseeable harms, algorithmic bias, and inadequate testing, as well as surreptitious or unexpected data collection, unmonitored data pipelines, and unvetted third-party use), and current proposals, such as the American Privacy Rights Act discussion draft and the Traveler Privacy Protection Act of 2025, signal a possible national baseline that makes transportable, PET-backed analytics a pragmatic default \cite{FTC_Biometric_Policy_2023,APRA_Sensitive_2024,TPPA_2025}.

In \Cref{sec:theory}, we discuss foundational theories that can advance the adoption of privacy-enhancing technologies in individualized and digitized healthcare. Combined with engineering designs that facilitate user experience and regulatory compliance in \Cref{sec:engineering}, these foundational theories help us progress toward a community of practice centered on health data privacy. In \cref{sec:applications}, we discuss use cases where developments in the responsible AI domain can significantly impact healthcare and raise awareness and appreciation of privacy-enhancing technologies and privacy-preserving analytics in health research and practice for regulatory and ethical compliance. We end with concluding remarks in \Cref{sec:conc} and describe the opportunities and challenges where theoretical and engineering advances can be most impactful.

\section{Foundations of Scalable Privacy in Networked Healthcare} \label{sec:theory}
%\textbf{From rows to flows.} 
\vspace{-3pt}
Real-world health data are \emph{interdependent}: patients share households and communities; providers collaborate in teams; sites contribute to multi-center studies; and public-health actions diffuse over contact and referral networks. In such settings, privacy failures propagate along \emph{connections}, not just through record release. Guarantees must therefore protect \emph{information flows} (exposures, neighborhood statistics, belief updates), not only static tables.

 Differential privacy (DP) offers a robust privacy definition that has emerged as a gold standard for privacy-preserving analytics to balance statistical precision and credibility with privacy risks to individuals. Conventional anonymization, hashing and encryption techniques can be tedious, ad hoc, and costly, and are more suitable for data curation among trusted parties because they have been shown to be vulnerable to various types of identification \cite{barbaro2006face}, linkage and cross-referencing \cite{sweeney2015only,mcsherry2007mechanism,sweeney1997weaving}, pseudo-reidentification \cite{hallaj_open_2025}, and statistical difference and reidentification attacks \cite{kumar2007anonymizing}; all together they fail to stand up to privacy challenges of digitized and individualized healthcare. 
 DP has been very influential in modern privacy research, including large-scale and high-profile applications in government, e.g., the US decennial census \cite{gong2022harnessing} and the tech industry (e.g., tracking statistics of emoji usage \cite{erlingsson2014learning,appleDP}), but practical and large-scale implementations of DP in healthcare remain limited (perhaps with the exception of the recently released microdata from Israel's National Registry of Live Birth \cite{hod2025differentially,hod2023designing}).

DP requires that the distribution of the output of an algorithm (mechanism) acting on data remain almost unchanged by the presence, absence, or modification of an individual’s record, thus limiting what an adversary (or any observer) can infer about the input by observing the output \cite{dwork2011firm,dwork14}. This is typically achieved through randomization in the algorithm design. For example, a randomization scheme can add noise to the original data in such a way that an adversary cannot tell whether or not any individual's data were included in the original data set. It gives a formal guarantee of privacy as follows: A mechanism $\cM: \cS \to \cR$ acting on patients data, $s \in \cS$, is $(\eps, \delta)$-DP if $\Pr [\cM(\vec s) \in \mathfrak R] \le e^\eps \Pr [\cM(\mathfrak \vec s') \in \mathfrak R] + \delta$ for any pair of adjacent observations $\vec s, \vec s'$ and any subsets $\mathfrak R$ of the range space $\cR$. The $(\eps, \delta)$-DP constraint on $\cM$ ensures that as $\eps,\delta \to 0$, no information can be inferred about whether $s$ or any of its adjacent points are the input generating observed output $\cM(\vec s)$. The adjacency notion between datasets is typically defined such that neighboring datasets differ in contributions of an individual entity, and thus $(\eps, \delta)$-DP protects individuals' contributions from ramification of an adversarial inference. Hence, different notions of adjacency will imply different protections for the individuals. This information-theoretic guarantee makes the algorithms output safe to release even against the threats of quantum computing (or any computationally unbounded adversary).  Parameter $\eps$ is referred to as privacy budget with $\eps \to \infty$ corresponding to the non-private case where no privacy protections are afforded. When $\delta = 0$ we recover the pure differential privacy constraint ($\eps$-DP) which is relaxed by $\delta>0$. As a concrete example in healthcare settings, consider the decision to participate in a clinical trail for a patient with a cardiovascular condition that puts them at a $2\%$ increased risk of stroke over the $0.01\%$ in general population. If known to insurers, the condition will increase patient's premium by $\$600$ to account for the $2\%$ increased chance of a stroke that would cost $\$30,000$. This can discourage patients from participating in the trial; however, if analyzed in an $\eps$-DP manner with $\eps=1$, the insurer's inferred risk by observing the trial data will increase by at most $0.01 (e^1-1)\% = 0.017\%$ corresponding to a $\$5.1$ increase in the premium cost, which may then be acceptable to the patient. Adding noise to the input data helps enforce the $(\eps,\delta)$-DP constraint in different settings but the added noise can also degrade performance, e.g., lowering the quality of distributed estimation \cite{papachristou_rahimian_dpdi_medrxiv_2025,papachristou2025differentially} or targeted interventions \cite{rahimian2023seeding,rahimian2023aamas}. The subsequent privacy-utility tradeoffs are complex to resolve and require careful modeling of the decision pipeline in healthcare settings. In conjunction with DP, contextual integrity can be a design lens \cite{nissenbaum04,benthall2024integrating} to adapt technical measures to clinical and public-health norms, with the goal of \emph{providing {auditable} guarantees at scale by treating privacy as a systems property that flows with data and decisions, not just with files}.

 {\bf State-of-the-art.} Current approaches to privacy-preserving analytics in healthcare settings are primarily application-specific, for example, in genomics \cite{wan2022sociotechnical}, for synthetic generation of electronic health records \cite{yoon2023ehr} and clinical trial data \cite{eckardt2024mimicking,beaulieu2019privacy}. Federated learning (FL) is a popular paradigm for decentralized optimization and distributed training of ML and AI models \cite{shi2023ensemble,yue2024federated} builds on the principle of data minimization to mitigate privacy barriers by eliminating the need to collect all data points on a central server for gradient calculations \cite{mcmahan2022federated,kontar2021internet}, with promising applications for predictive analytics in healthcare \cite{kaissis2020secure,xu2021federated}. While FL provides basic privacy protection, depending on different threat assumptions, various privacy concerns still exist and various privacy-preserving FL approaches have been proposed in the literature \cite{bonawitz2022federated}.  %, wireless communications \cite{niknam2020federated}, etc. 
 Differential privacy (DP) is another popular approach that limits privacy leakage by injecting controlled noise and randomness into the algorithm output \cite{mcsherry2007mechanism,zhu2017differentially}, and has been successfully applied to biomedical data analysis (e.g., medical imaging \cite{ziller2021medical} and genomics data privacy \cite{halimi2022privacy}) and sharing \cite{yamamoto2021more,uhler2013privgenomics}, including in conjunction with FL \cite{adnan2022scirep,choudhury2019differential}. Last on the list, cryptographic techniques such as homomorphic and functional encryption and multiparty computing allow computations on encrypted data with promising applications in healthcare settings \cite{blatt2020secure,geva2023collaborative}; however, the strong cryptographic guarantees of these methods come at a high cost to computational efficiency, and their implementations are limited by information technology (IT) requirements and computing resources that are beyond the technical capabilities of many organizations and devices. Trusted execution environments or more generally now known as confidential computing provides a faster alternative to and complements other approaches via a hard-ware-rooted chain-of-trust for end-to-end privacy solutions \cite{khan2025privacy,geppert2022trusted,paju2023sok}.
%There are also DP prediction problems, where distributed agents share data to make common predictions based on statistical properties of the distributed data (e.g. mean or maximum)  \cite{yang2023local}. 
%In contrast, \textit{we adopt a decision-theoretic framework in which healthcare operations are informed by privatized data, but the collective goal (e.g., maximizing public impact of an intervention or establishing collective evidence of superiority of a new intervention with as few samples as possible) is more complex than ensuring statistical accuracy of an aggregate or predictive performance of an ML algorithm in typical applications of differential privacy and federated learning.} 
% y faster runtime than existing encryption-based privacy-enhancing technologies (PETs) and federated analytics for multicenter
% studies (Froelicher et al. 2021). state-of-the-art distributed survival analysis methods based on homomorphic
% encryption (Froelicher et al. 2021, Geva et al. 2023).  secure multiparty computation (MPC) and homomorphic encryption (HE) 

{\textit{What is lacking is a framework that can be applied flexibly and consistently to healthcare delivery and public health operations, that is expressive enough to allow us to balance the risks and benefits of data sharing for healthcare consumers and their communities, and that goes beyond typical criteria such as statistical accuracy of estimates or predictive performance of AI and ML models in state-of-the-art adaptations of DP and FL to healthcare settings}.

\textbf{Decision-theoretic DP for privacy from data to decisions.} DP can be embedded as statistical disclosure control throughout the data and model lifecycles such that guarantees align with \emph{decisions}---e.g., trial enrollment, resource allocation, and targeting interventions---and \emph{harms}---e.g., insurance rate hikes and stigmatization. For trials, DP can limit inferences that an adversary could draw from aggregate signals while preserving statistical power; for disease surveillance, DP can stabilize risk communication without exposing individuals. This redefines privacy as a \emph{design variable} to be optimized alongside statistical power, fairness, accuracy, and cost \cite{ziller2021medical,rogers2016privacy}.

\textbf{Network-aware privacy for interdependent data.} Privacy mechanisms should protect \emph{individuals and ties} as information diffuses in networks, with sensitivity that scales with network topology (degree distribution, clustering, and path length) and dynamics (sequential learning and feedback) among networked agents. Network-aware privacy guarantees can support applications such as \emph{distributed trials}, \emph{biomarker discovery}, and \emph{network interventions} while limiting group-level information leakage and cumulative participation risks \cite{kasivis2013nodeDP,hay2009graphdp}. Foundational progress in network-aware privacy can address challenges that arise when data points are interdependent, as in social and biomedical networks, and offer solutions that protect privacy in information flows rather than static network structures, allowing rigorous analysis of privacy and utility in dynamic settings such as contact tracing, biomarker discovery, distributed trials, and network-based interventions.

{\bf Distributed inference as a workhorse.} Many networked tasks reduce to \emph{distributed evidence accumulation}: sites compute local statistics and exchange privatized \emph{belief updates}; a network of sites reaches decisions (e.g., superiority in a trial, change-point alarms). Injecting DP noise into log-likelihoods or other sufficient statistics protects \emph{what the network learns} about any contributor while preserving decision utility. We refer to \emph{Differentially Private Distributed Inference} (DPDI) as a framework for multi-site learning and sequential updates under privacy constraints \cite{papachristou_rahimian_dpdi_medrxiv_2025}.

\section{Engineering Scalable Privacy in Networked healthcare Systems}\label{sec:engineering}

%\textbf{Limits of point solutions.} 
Privacy-enhancing technologies (PETs) are powerful but fragmented. Synthetic electronic health records (EHRs) aid dissemination but require risk controls to avoid leakage \cite{yoon2023ehr,tian2024reliable}. Federated learning (FL) reduces central aggregation, yet model updates leak information unless paired with formal privacy \cite{truex2019hybrid,bonawitz2019towards,mcmahan2017communication}, and clinical deployments increasingly combine FL with DP \cite{adnan2022scirep}. %DP for imaging and healthcare \cite{ziller2021medical,ficek21,khalid23}.
Cryptography techniques, such as secure multi-party computation (MPC) and homomorphic encryption (HE), protect confidentiality but may be computationally heavy; recent systems, e.g., federated analytics based on multiparty homomorphic encryption (FAMHE), show feasibility for cross-silo analytics \cite{froelicher21}. In practice, deployments mix these tools—often without a unifying privacy budget or end-to-end accounting.

Prior work has adapted differential privacy to medical imaging and EHR prediction, yielding task-useful models with record-level guarantees in centralized and federated settings \cite{ziller2021medical,ficek2021differential,khalid2023privacy}. These systems typically assume i.i.d. data and point solutions that involve single-task training and report utility under one-shot privacy accounting. What remains under-specified is how privacy risk composes over repeated participation, across institutions, and along patient or household networks. In this section we offer design principles for allocating privacy budgets and certifying leakage in such networked, multi-institution contexts.
%\textcolor{red}{merge and remove} \textbf{PETs in health.} %FL in medicine (surveys and applications) \cite{guan24},  DP for imaging and healthcare \cite{ziller2021medical,ficek21,khalid23}, %and HE/MPC for cross-silo analytics \cite{froelicher21,truhn24} demonstrate feasibility and limits. %Synthetic EHR methods yield shareable artifacts with bounded risks \cite{yoon2023ehr,tian24}.
Health systems learn continuously, exchange models across institutions, and reuse representations downstream; under these operations, risk compounds and can spill over along family, provider, or geographic networks. We treat privacy as a \emph{systems} property spanning (i) \textbf{data collection} (edge instrumentation, consent, sampling, logging), (ii) \textbf{inference} (statistical and machine learning modeling pipelines), and (iii) \textbf{actions} (clinical, operational, public-health decisions). To that end, we envision engineering advances that move from “{DP per model}” to “{DP across the learning workflow},” by \textit{(i) explicit privacy budgeting over time and across sites, (ii) network-aware guarantees that bound group- and linkage-level leakage, and (iii) auditable composition integrated with compliance-as-code}. %with explicit budgeting, composition, and network-aware guarantees. We treat privacy as a \emph{systems} property spanning (i) \textbf{collection} (edge instrumentation, consent, sampling, logging), (ii) \textbf{inference} (stat/ML pipelines), and (iii) \textbf{action} (clinical, operational, public-health decisions).

%Prior work shows that useful medical imaging and EHR models can be trained under record-level DP in centralized and federated settings \cite{ziller2021medical,ficek21,khalid23}. But most pipelines target single tasks and populations, assume roughly i.i.d. data, and report utility under one-shot accounting. Learning health systems, by contrast, compose risk: models are retrained and reused across sites and time; representations flow downstream; correlations link patients, households, providers, and institutions. %Our agenda moves from “DP per model” to “DP across the workflow” with (i) explicit budgeting over time and across sites, (ii) network-aware guarantees that bound group- and linkage-level leakage, and (iii) auditable composition integrated with compliance-as-code. We treat privacy as a systems property spanning (a) collection (edge instrumentation, consent, sampling, logging), (b) inference (stat/ML pipelines), and (c) action (clinical, operational, public-health decisions).

\textbf{Data models and interoperability.} The Observational Medical Outcomes Partnership (OMOP), Observational Health Data Sciences and Informatics (OHDSI) and the national patient-centered clinical research network (PCORnet) common data models (CDMs) standardize schema and vocabularies for reproducible, distributed analytics in healthcare research \cite{ohdsi20,pcornet25}. For example, the Artificial Intelligence-Ready and Equitable Atlas for Diabetes Insights (AI-READI) project structures the AI-READI dataset using the OMOP CDM for clinical data and DICOM format for images \cite{owsley_cross-sectional_2025}. However, the AI-READI project built a project-specific platform for sharing purposes (\href{fairhub.io}{fairhub.io}) with a bespoke, multi-layered ``Swiss-cheese" access model to manage privacy risks \cite{hallaj_open_2025}. To scale safely, we need PETs that can integrate with clinical data models (OMOP/PCORnet) and produce verifiable compliance evidence; i.e., PET components that integrate with these data ecosystems with pre-audited queries and provenance, instead of forcing major research consortia to build their own custom governance platforms \cite{hallaj_open_2025}.

\textbf{Compliance-as-code and the future of PETs.} Machine-readable policies, {privacy budget ledgers}, data lineage, access logs, and auto-generated IRB/DUA texts can help operationalize compliance-as-code. To increase interoperability, common statistical tasks such as survival analysis can be mapped to pre-audited queries against common data models \cite{hripcsak2016characterizing,fleurence2014launching} and align with OMOP/PCORnet schemas for plug-in adoption. NIST Privacy Framework and AI RMF \cite{nistpf20,nistairmf23} can inform integration of such standardized queries within organizational risk management frameworks. Future PETs can combine \emph{federated analytics} to keep raw data in place, use \emph{DP} to cap leakage from information flow between updates and releases, and use \emph{MPC/HE} for high-risk subroutines that involve un-trusted aggregator and cross-site linkage \cite{froelicher21}, and offer modular proofs to support end-to-end privacy certifications.

\section{Use Cases, Testbed and Metrics for a Use-Inspired Agenda}\label{sec:applications}
% {\bf Genomics and High-Dimensional Biomarkers}
% Population-scale studies need strong controls against membership and linkage attacks. DP for feature selection and model release, paired with access brokering and post-processing checks, provides \emph{bounded-risk} sharing while preserving reproducibility \cite{khalid23,uhler_privgenomics_2013,shin_priv_gwas_2017}. \textcolor{red}{add from proposal}

\subsection{Multi-omics and High-Dimensional Biomarkers}

Genomics, proteomics, and other high-dimensional biomarkers are becoming routine in learning health systems. Population-scale studies need strong controls against membership and linkage attacks. While, DP for feature selection and model release, paired with access brokering and post-processing checks, provides controls against membership and linkage attacks in population-scale studies \cite{khalid2023privacy,uhler2013privgenomics}, the scale and granularity of  genome-wide association study (GWAS)-style settings amplify familiar membership and linkage attack risks and raise a less discussed class of \emph{socio-biomedical} spillovers in which data about one person can reveal private facts about relatives, household members, or contacts \cite{homer2008resolving,Gymrek2013Science,wan2022sociotechnical}. As cohorts diversify and data moves more fluidly across institutions, treating such relationships as first-class considerations becomes essential.

What would privacy look like if we designed \emph{for} these realities from the start? One opportunity is to view biomarker discovery not as a single release but as a \emph{workflow} in which selection, testing, sharing, and reuse are visible to governance. In that view, differential privacy offers guardrails for per-record disclosure; federated and encrypted analytics reduce needless aggregation; and policy-aware audit trails help make reuse accountable \cite{dwork14,pulivarti2024genomic}. The open question is how these ingredients might cohere so that privacy “travels with” biomarkers—across sites, time, and downstream applications—without derailing scientific utility. 

We see several directions worth exploring:
\begin{itemize}
  \item \textbf{Relationship-aware sensitivity.} Many biomarker tasks have natural structure (kinship, shared exposure, phylogeny). Accounting for that structure when bounding sensitivity could reduce worst-case leakage while preserving power on signals that practitioners care about \cite{Yilmaz2022CODASPY,Almadhoun2020Bioinformatics}.
  \item \textbf{Selection under uncertainty.} Discovery pipelines frequently perform top-$k$ selection of variants, regions, or panels. Lightweight mechanisms for private selection—paired with simple, shareable summaries—may strike a practical balance between transparency and risk \cite{Qiao2021ICML,mcsherry2007mechanism}.
  \item \textbf{Privacy that composes over time.} Health systems learn continuously. Rather than a one-shot guarantee, budgeting and provenance could make it easier to understand how risk accumulates across iterations and across institutions \cite{dwork14,pulivarti2024genomic}.
  \item \textbf{Stress-testing as a norm.} Beyond accuracy, routine checks for membership inference, kinship recovery, or contact-path inference could become part of method validation—akin to robustness checks in other domains \cite{homer2008resolving,Gymrek2013Science}.
\end{itemize}

None of these ideas require a monolithic solution. A more realistic near-term picture is a set of interoperable practices: modest noise when it helps, pre-audited queries where they fit, and clear records of what was done and why. If adopted consistently, such practices could make genomics-driven learning both \emph{useful} and \emph{appropriately constrained}, even as datasets, models, and collaborations grow.

\subsection{Distributed Clinical Trials and Multi-Site Learning}
The need for scalable privacy frameworks that can federate analysis across institutions while protecting sensitive demographic and clinical data is demonstrated by multi-site efforts like the AI-READI project, which is harmonizing data from three distinct geographical sites and triple-balancing by race/ethnicity, sex, and diabetes status for training ``unbiased machine learning models" \cite{owsley_cross-sectional_2025}. Federated trial analytics with DP can accelerate recruitment, subgroup analyses, and safety monitoring, while providing cross-site auditability. Imaging and pathology studies illustrate practicality of combining federated learning and DP \cite{adnan2022scirep} and encrypted analytics demonstrate viability for survival analysis and GWAS-like tasks \cite{froelicher21}. 

  Clinical trials are the gold standard for generating medical evidence and evaluating interventions. When testing treatments for emerging diseases, recruiting patients for clinical trials is challenging, especially if the disease is rare and/or severe and there are not many drugs available. One solution is to establish a central network of clinical trials. In the case of AIDS this was achieved through the AIDS Clinical Trials Group (ACTG) network, which required a huge prior investment and central coordination. However, more generally, privacy considerations, often in the form of institutional and regulatory hurdles, hamper data sharing in such settings. For example, multicenter clinical trials typically require the execution of extensive data usage agreements that slow down or prevent collaborations. The differentially private distributed inference (DPDI) framework operationalizes privacy-preserving hypothesis testing and survival analysis across sites by exchanging only DP-sanitized belief statistics, reducing reliance on complex DUAs while remaining compatible with HIPAA/GDPR constraints and site governance \cite{papachristou_rahimian_dpdi_medrxiv_2025}. This aligns with the ARPA–H Advancing Clinical Trial Readiness (ACTR) vision for decentralized, on-demand trials and automated EHR (Electronic health Record)-Case Report Form (CRF) synchronization. DPDI enables sites to contribute analysis-ready statistics without exporting raw PHI, supporting recruitment, subgroup analyses, and safety monitoring at scale. In contrast to cryptography-heavy PETs (HE/MPC), which provide strong formal guarantees but often demand substantial IT/HPC capacity and specialized engineering across sites, DPDI achieves competitive accuracy–privacy tradeoffs with far lower computational and orchestration overheads \cite{froelicher21,geva2023collaborative}. Consistent with emerging open-science practices (e.g., AI-READI’s governance for multi-site, balanced datasets), DPDI gives auditable pipelines that harmonize with federated analytics in healthcare \cite{hallaj_open_2025,bonawitz2022federated,ficek2021differential}.

%\noindent\textit{Adaptive recruitment and retention under DP.}
Beyond fixed trial analytics, network-aware DP seeding can drive \emph{adaptive} site- and participant-selection policies that improve power and subgroup coverage while respecting privacy budgets. In this view, interim analyses expose only privatized sufficient statistics (e.g., DP scores and log-likelihood increments), and recruitment and retention policies can prioritize high-value strata or sites subject to fairness and leakage constraints. The seeding-with-DP framework provides the targeting mechanism from partial contact and referral traces \cite{rahimian2023seeding}, while DPDI supplies the privacy-preserving evidence-accumulation layer for sequential monitoring and deciding superiority or futility of treatment arms \cite{papachristou_rahimian_dpdi_medrxiv_2025}. These components enable a closed-loop, data-to-decision pipeline in which (i) who to invite next, (ii) which sites to emphasize, and (iii) when to stop for success or futility are all tuned to formal privacy budgets and audit-ready composition accounting.

\subsection{Public health operations: Contact tracing, disease surveillance, targeted screening and preventive measures}
%\vspace{-5pt}

The COVID-19 pandemic brought the public health machinery to the forefront of public discourse. Memories of nonpharmaceutical interventions, contact tracing, quarantine operations, and vaccination rollouts are still fresh in many minds \cite{holtz2020interdependence,collis2022global}. A functioning public health system is essential to the health and safety of our nation. This role comes into the spotlight with every emerging infectious disease (Legionnaires' disease in the 1970s, AIDS in the 1980s, and SARS and its variants in the 2000s). Outside the spotlight, public health function remains essential in the management of endemic diseases such as influenza in all flu seasons and long-lasting issues such as the US opioid epidemic. The more than 200-year history of public health services in the US has produced many success stories in fighting for the lives and livelihoods of its people. The eradication of smallpox \cite{henderson2011eradication} and the accomplishments of the PEPFAR initiative in preventing HIV infection and AIDS \cite{reid2024programme} are two examples from recent history. However, this very public function comes with significant challenges. During the COVID-19 pandemic, some of the potentially broad public health impacts were undermined by dysfunctions within the information ecosystems of democracies manifested in widespread misinformation about vaccination \cite{cinelli2020covid,loomba2021measuring,skafle2022misinformation} and the erosion of trust between people, public health agencies, and the scientific community \cite{nahum2021erosion}. Complicating the challenges of broken trust levels is the increasing role of AI and data-driven frameworks in public health that come with great potential to improve operational efficiency, but also risks exacerbating existing inequalities and worsening the distrust of public health apparatus when threatened with individual privacy violations and faced with digital technologies that lack transparency and accountability in dealing with people's lives and livelihoods. The dangers in this case are not limited to trust erosion and can take other forms, for example, harassment targeted at public officials and the scientific community, as observed with the anti-vaccine movement \cite{carpiano2023confronting,rahimian2025democratic}.

Exposure notification, sentinel placement, and outbreak detection are network optimization problems under constraints. Network-aware DP combined with network optimization frameworks such as submodular sensor selection can preserve effectiveness while bounding disclosure risk. Contact tracing routinely yields \emph{influence samples} which are partial cascades and short contact paths, rather than a full network. Leveraging such samples for optimizing sentinel surveillance, targeting, or seeding (e.g., selecting initial recipients for scarce prevention resources) creates a dual requirement: (i) learn enough structure from partial data to achieve high intervention impact, and (ii) protect the privacy of individuals and ties revealed by tracing. Recent work formalizes this setting and introduces privacy notions and algorithms for \emph{seeding with differentially private network information} when only influence samples are available \cite{rahimian2023seeding,rahimian2023aamas}. 

Two deployment regimes arise for DP. In \emph{central DP (CDP)}, a trusted curator observes raw contact samples but releases a randomized seed set; in \emph{local DP (LDP)}, randomization is applied at data collection so that the aggregator never sees unprotected samples. Theory and simulations show graceful performance degradation with decreasing privacy budgets in CDP and clarify the (higher) budget requirements for strong LDP guaranties \cite{rahimian2023seeding}. This framework is directly applicable to HIV prevention with pre-exposure prophylaxis (PrEP), where seeding high-impact individuals in sexual contact networks of men who have sex with men (MSM) can amplify population-level benefits. Empirically grounded network models from ARTnet provide realistic degree and mixing patterns for MSM in the U.S. \cite{weiss2020artnet}, while health-economic studies consistently support prioritizing higher-risk subpopulations to improve the cost-effectiveness of PrEP \cite{juusola2012prep,kessler2014prep}. A \emph{network-aware, DP} targeting pipeline thus permits: (a) extracting noisy, privacy-preserving influence samples from tracing and surveillance workflows; (b) selecting targets that respect formal privacy budgets for individuals and ties; and (c) quantifying privacy–utility trade-offs for operational policies (e.g., number of targets, cadence, and re-targeting rules). Beyond PrEP, the same approach extends to other outbreak responses (e.g., exposure notification and sentinel placement), where the objective is early detection or transmission suppression under rigorous privacy accounting.

\subsection{Mobile  Health (mHealth) Ecosystems}

Systematic reviews point to gaps between regulatory intent and app practice. Privacy-by-design stacks (DP queries, FL with secure aggregation, data enclaves and cryptographic techniques) plus tamper-evident logs can align telemetry with system-level assurance and clinician review \cite{ficek2021differential}. This is essential for modern study protocols that deploy mHealth devices at scale, such as the AI-READI study's use of 10-day continuous glucose monitors (CGMs) and Garmin physical activity trackers for roughly 4,000 participants \cite{owsley_cross-sectional_2025}. Proper orchestration of PETs can provide the mHealth privacy ``operating system" to support processing of high-velocity telemetry data.

% \section{Tools, Testbeds, and Metrics (Practice-Oriented Agenda)}
% \textcolor{red}{merge and remove} 
\section{Conclusions, Opportunities and Challenges}\label{sec:conc}

From continuous biometrics and wearables to multi-omics and learning health systems, healthcare is increasingly \emph{smart}, \emph{networked}, and \emph{AI-enabled}. These advances depend on data flows across clinics, labs, payers, vendors, and public-health agencies---and ultimately on public trust that sharing data will not cause harm \cite{kuan2019adopting,wang2022privacy}. Governance scaffolding (GDPR, HIPAA, GINA, and the Common Rule) sets the floor, while the NIH Data Management and Sharing (DMS) Policy and the NIST Privacy Framework and AI Risk Management Framework push toward broader use under stronger safeguards \cite{hipaa_privacy_rule,gina_2008,common_rule_2018,nihdms23,nistpf20,nistairmf23,gdpr_text_2016}. Yet the speed, granularity, and context-dependence of modern data use demand frameworks that reason about \emph{privacy--utility trade-offs} in situ, not merely the post-hoc de-identification model \cite{nass2009value,sweeney1997weaving,sweeney2015only} that is increasingly challenged by new attack vectors, such as ``pseudo-reidentification", where personally identifiable information is not needed to single out unique individuals in datasets containing wearable, electrocardiogram, or retinal imaging data \cite{hallaj_open_2025}.

Realizing privacy at scale in networked healthcare poses several interrelated challenges in practice: (i) rigorous, end-to-end privacy accounting and composition across sites, datasets, and successive analyses, including in federated settings \cite{mcmahan2022federated,kairouz21}; (ii) robustness to modern attacks and realistic threat models for clinical machine learning pipelines \cite{tramer2020adaptive}; (iii) data harmonization and equity under heterogeneity by linking EHR, medical images, and other healthcare data modalities such that that protections do not disproportionately degrade utility for minority subgroups \cite{ziller2021medical,yoon2023ehr}; (iv) governance, auditability, and provenance for federated analytics to align technical assurances with emerging organizational frameworks \cite{nistpf20,nistairmf23,hipaa_privacy_rule,NIH_DMS_2021,nihdms23}; (v) practical mechanisms for consent management, withdrawal, and machine unlearning in adaptive and streaming studies \cite{neel2021descent,ginart2019making,BIPA_SB2979_2024}; (vi) extending protections beyond individuals to families, communities, and contact networks in socio-biomedical contexts \cite{wan2022sociotechnical,nissenbaum04}; (vii) engineering trade-offs among PETs (including differential privacy, secure aggregation, multiparty computation, homomorphic encryption, trusted execution environments, and synthetic data techniques) to meet accuracy, latency, and cost constraints in real workflows \cite{bonawitz2017secureagg,bonawitz2019towards,froelicher21,geva2023collaborative,blatt2020secure,yue2024federated}; and (viii) incentive-compatible sharing and evaluation with public benchmarks and cross-site audits that align institutional risk, regulatory change, and scientific value \cite{weiss2020artnet,NIH_DMS_Reminder_2023}. 

To address these challenges, we envision engineering stacks that make privacy an auditable, first-class system property. For example, design wizards can translate privacy regulations and policy clauses, such as secondary use, limited data set, and time-bound retention, into flow diagrams, PET compositions, and DUA-ready text with traceable rationales; and privacy-budget ledgers can facilitate differential privacy implementations with calculators, composition tracking, per-dataset “privacy labels,” and pipeline APIs that support governance across projects and sites (drawing lessons from the Census TopDown deployment’s centralized allocation, cross-table composition, and public documentation \cite{abowd_censusdp_2019}). Differential privacy primitives in distributed inference can facilitate information sharing and learning across institutions with privatized belief updates and hypothesis-test routines \cite{papachristou_rahimian_dpdi_medrxiv_2025}. A PET coordination layer can provide a control plane to combine DP, secure aggregation, homomorphic encryption, and MPC to enforce data use purpose, recipient, and retention guards, capture provenance, and bundle reusable audit artifacts with proofs bound to machine-readable configurations \cite{blatt2020secure,geva2023collaborative,bonawitz2017secureagg}. Finally, compliance sandboxes and curated, consented, and synthetic testbeds can provide attacker models and red-teaming capabilities that generate legible risk evidence for realistic networked tasks. For example, one can use high-fidelity synthetic EHR resources to devise testbeds in patient referral networks or multi-site trials \cite{yoon2023ehr,tian2024reliable}. Together, these pieces align with the national emphasis on usable tools, integrated solutions, and testbeds for privacy-preserving data sharing in practice \cite{pdasp24}. Progress on these fronts will enable privacy-preserving and learning healthcare systems that are ethical, equitable, auditable, and deployable in routine care.

% \section{Related Work}
% \textbf{Foundations and open problems.} DP foundations and open problems ground privacy accounting and utility analysis \cite{dwork14,kairouz21}. 

% \textbf{Data models and interoperability.} OMOP/OHDSI and PCORnet CDMs standardize schemas and vocabularies for reproducible, distributed analytics \cite{ohdsi20,pcornet25}. Our agenda targets PET components that ``snap in'' to these ecosystems with pre-audited queries and provenance.

% \textbf{Policy frameworks.} NIH DMS and NIST Privacy/AI RMFs inform governance and risk management, motivating \emph{compliance-as-code} artifacts in pipelines \cite{nihdms23,nistpf20,nistairmf23}.

% \section{Open Problems}

\bibliographystyle{IEEEtran}
\bibliography{privacy_networked_healthcare}

@IEEEtranBSTCTL{IEEEexample:BSTcontrol,
  CTLuse_doi = "no",
  CTLuse_url = "no",
  CTLuse_eprint = "no",
}

@misc{hipaa_privacy_rule,
  title        = {Standards for Privacy of Individually Identifiable Health Information (HIPAA Privacy Rule)},
  howpublished = {\url{https://www.hhs.gov/hipaa/for-professionals/privacy/index.html}},
  year         = {2002},
  note         = {{US} Department of Health and Human Services}
}

@misc{gina_2008,
  title        = {Genetic Information Nondiscrimination Act of 2008 ({GINA}), Public Law 110–233},
  howpublished = {\url{https://www.eeoc.gov/statutes/genetic-information-nondiscrimination-act-2008}},
  year         = {2008}
}

@misc{common_rule_2018,
  title        = {Federal Policy for the Protection of Human Subjects (Common Rule), 45 {CFR} 46 (2018 Requirements)},
  howpublished = {\url{https://www.hhs.gov/ohrp/regulations-and-policy/regulations/45-cfr-46/index.html}},
  year         = {2018}
}

@misc{nihdms23,
  title        = {{NIH} Policy for Data Management and Sharing},
  howpublished = {\url{https://sharing.nih.gov/data-management-and-sharing-policy}},
  year         = {2023}
}

@misc{nistpf20,
  title        = {{NIST} Privacy Framework: A Tool for Improving Privacy through Enterprise Risk Management},
  howpublished = {\url{https://www.nist.gov/privacy-framework}},
  year         = {2020}
}

@misc{nistairmf23,
  title        = {{NIST AI Risk Management Framework (AI RMF 1.0)}},
  howpublished = {\url{https://www.nist.gov/itl/ai-risk-management-framework}},
  year         = {2023}
}

@techreport{abowd_censusdp_2019,
  author       = {John Abowd and Robert Ashmead and Simson Garfinkel and Daniel Kifer and Philip Leclerc and Ashwin Machanavajjhala and William Sexton},
  title        = {Census {TopDown}: Differentially Private Data, Incremental Schemas, and Consistency with Public Knowledge},
  institution  = {US Census Bureau},
  year         = {2019}
}

@misc{gdpr_text_2016,
  author       = {{European Parliament and Council of the European Union}},
  title        = {Regulation ({EU}) 2016/679 (General Data Protection Regulation)},
  howpublished = {\url{https://data.europa.eu/eli/reg/2016/679/oj}},
  year         = {2016}
}

@article{nissenbaum04,
  author  = {Helen Nissenbaum},
  title   = {Privacy as Contextual Integrity},
  journal = {Washington Law Review},
  volume  = {79},
  pages   = {119--158},
  year    = {2004}
}

@article{wan2022sociotechnical,
  author  = {Zhiyu Wan and James W. Hazel and Ellen W. Clayton and Yevgeniy Vorobeychik and Murat Kantarcioglu and Bradley A. Malin},
  title   = {Sociotechnical safeguards for genomic data privacy},
  journal = {Nature Reviews Genetics},
  volume  = {23},
  number  = {7},
  pages   = {429--445},
  year    = {2022}
}

@article{yoon2023ehr,
  author  = {Jinsung Yoon and Michel Mizrahi and Nahid Farhady Ghalaty and Thomas Jarvinen and Ashwin S. Ravi and Peter Brune and Fanyu Kong and Dave Anderson and George Lee and Arie Meir and others},
  title   = {{EHR}-Safe: Generating high-fidelity and privacy-preserving synthetic electronic health records},
  journal = {NPJ Digital Medicine},
  volume  = {6},
  number  = {1},
  pages   = {141},
  year    = {2023}
}

@article{eckardt2024mimicking,
  author  = {Jan-Niklas Eckardt and Waldemar Hahn and Christoph R{\"o}llig and Sebastian Stasik and Uwe Platzbecker and Carsten M{\"u}ller-Tidow and Hubert Serve and Claudia D. Baldus and Christoph Schliemann and Kerstin Sch{\"a}fer-Eckart and others},
  title   = {Mimicking clinical trials with synthetic acute myeloid leukemia patients using generative artificial intelligence},
  journal = {NPJ Digital Medicine},
  volume  = {7},
  number  = {1},
  pages   = {76},
  year    = {2024}
}

@article{beaulieu2019privacy,
  author  = {Brett K. Beaulieu-Jones and Zhiwei Steven Wu and Chris Williams and Ran Lee and Sanjeev P. Bhavnani and James Brian Byrd and Casey S. Greene},
  title   = {Privacy-preserving generative deep neural networks support clinical data sharing},
  journal = {Circulation: Cardiovascular Quality and Outcomes},
  volume  = {12},
  number  = {7},
  pages   = {e005122},
  year    = {2019}
}

@inproceedings{halimi2022privacy,
  title={Privacy-preserving and efficient verification of the outcome in genome-wide association studies},
  author={Halimi, Anisa and Dervishi, Leonard and Ayday, Erman and Pyrgelis, Apostolos and Troncoso-Pastoriza, Juan Ram{\'o}n and Hubaux, Jean-Pierre and Jiang, Xiaoqian and Vaidya, Jaideep},
  booktitle={Proceedings on Privacy Enhancing Technologies. Privacy Enhancing Technologies Symposium},
  volume={2022},
  number={3},
  pages={732--753},
  year={2022}
}

@article{yamamoto2021more,
  title={More practical differentially private publication of key statistics in {GWAS}},
  author={Yamamoto, Akito and Shibuya, Tetsuo},
  journal={Bioinformatics Advances},
  volume={1},
  number={1},
  pages={vbab004},
  year={2021},
  publisher={Oxford University Press}
}

@article{bonawitz2022federated,
  title={Federated learning and privacy},
  author={Bonawitz, Kallista and Kairouz, Peter and {McMahan}, Brendan and Ramage, Daniel},
  journal={Communications of the ACM},
  volume={65},
  number={4},
  pages={90--97},
  year={2022},
  publisher={ACM New York, NY, USA}
}

@article{shi2023ensemble,
  author  = {Naichen Shi and Fan Lai and Raed Al Kontar and Mosharaf Chowdhury},
  title   = {Ensemble models in federated learning for improved generalization and uncertainty quantification},
  journal = {IEEE Transactions on Automation Science and Engineering},
  year    = {2023}
}

@article{yue2024federated,
  title={Federated data analytics: A study on linear models},
  author={Yue, Xubo and Kontar, Raed Al and Gomez, Ana Maria Estrada},
  journal={IISE Transactions},
  volume={56},
  number={1},
  pages={16--28},
  year={2024},
  publisher={Taylor \& Francis}
}

@misc{mcmahan2022federated,
  author       = {Brendan {McMahan} and Abhradeep Thakurta},
  title        = {Federated Learning with Formal Differential Privacy Guarantees},
  howpublished = {Google {AI} Blog},
  year         = {2022},
  note         = {\url{https://ai.googleblog.com}}
}

@article{kontar2021internet,
  author  = {Raed Al Kontar and Naichen Shi and Xubo Yue and Seokhyun Chung and Eunshin Byon and Mosharaf Chowdhury and Jionghua Jin and Wissam Kontar and Neda Masoud and Maher Nouiehed and others},
  title   = {The Internet of Federated Things ({IoFT})},
  journal = {IEEE Access},
  volume  = {9},
  pages   = {156071--156113},
  year    = {2021}
}

@article{kaissis2020secure,
  author  = {Georgios A. Kaissis and Marcus R. Makowski and Daniel R{\"u}ckert and Rickmer F. Braren},
  title   = {Secure, Privacy-Preserving and Federated Machine Learning in Medical Imaging},
  journal = {Nature Machine Intelligence},
  volume  = {2},
  number  = {6},
  pages   = {305--311},
  year    = {2020}
}

@article{xu2021federated,
  author  = {Jie Xu and Benjamin S. Glicksberg and Chang Su and Peter Walker and Jiang Bian and Fei Wang},
  title   = {Federated Learning for Healthcare Informatics},
  journal = {Journal of Healthcare Informatics Research},
  volume  = {5},
  number  = {1},
  pages   = {1--19},
  year    = {2021}
}

@inproceedings{mcsherry2007mechanism,
  author    = {Frank McSherry and Kunal Talwar},
  title     = {Mechanism Design via Differential Privacy},
  booktitle = {48th Annual IEEE Symposium on Foundations of Computer Science (FOCS)},
  pages     = {94--103},
  year      = {2007},
  publisher = {IEEE}
}

@article{zhu2017differentially,
  author  = {Tianqing Zhu and Gang Li and Wanlei Zhou and  Philip S. Yu},
  title   = {Differentially Private Data Publishing and Analysis: A Survey},
  journal = {IEEE Transactions on Knowledge and Data Engineering},
  volume  = {29},
  number  = {8},
  pages   = {1619--1638},
  year    = {2017}
}

@article{ziller2021medical,
  author  = {Alexander Ziller and Dmitrii Usynin and Rickmer Braren and Marcus Makowski and Daniel Rueckert and Georgios Kaissis},
  title   = {Medical Imaging Deep Learning with Differential Privacy},
  journal = {Scientific Reports},
  volume  = {11},
  number  = {1},
  pages   = {13524},
  year    = {2021}
}

@misc{choudhury2019differential,
  author       = {Olivia Choudhury and Aris Gkoulalas-Divanis and Theodoros Salonidis and Issa Sylla and Yoonyoung Park and Grace Hsu and Amar Das},
  title        = {Differential Privacy-Enabled Federated Learning for Sensitive Health Data},
  howpublished = {NeurIPS ML4H Workshop; arXiv:1910.02578},
  year         = {2019}
}

@article{blatt2020secure,
  author  = {Marcelo Blatt and Alexander Gusev and Yuriy Polyakov and Shafi Goldwasser},
  title   = {Secure Large-Scale Genome-Wide Association Studies Using Homomorphic Encryption},
  journal = {Proceedings of the National Academy of Sciences},
  volume  = {117},
  number  = {21},
  pages   = {11608--11613},
  year    = {2020}
}

@article{geva2023collaborative,
  author  = {Ravit Geva and Alexander Gusev and Yuriy Polyakov and Lior Liram and Oded Rosolio and Andreea Alexandru and Nicholas Genise and Marcelo Blatt and Zohar Duchin and Barliz Waissengrin and others},
  title   = {Collaborative Privacy-Preserving Analysis of Oncological Data Using Multiparty Homomorphic Encryption},
  journal = {Proceedings of the National Academy of Sciences},
  volume  = {120},
  number  = {33},
  pages   = {e2304415120},
  year    = {2023}
}

@article{hripcsak2016characterizing,
  title={Characterizing treatment pathways at scale using the OHDSI network},
  author={Hripcsak, George and Ryan, Patrick B and Duke, Jon D and Shah, Nigam H and Park, Rae Woong and Huser, Vojtech and Suchard, Marc A and Schuemie, Martijn J and DeFalco, Frank J and Perotte, Adler and others},
  journal={Proceedings of the National Academy of Sciences},
  volume={113},
  number={27},
  pages={7329--7336},
  year={2016},
  publisher={National Academy of Sciences}
}

@article{fleurence2014launching,
  title={Launching PCORnet, a national patient-centered clinical research network},
  author={Fleurence, Rachael L and Curtis, Lesley H and Califf, Robert M and Platt, Richard and Selby, Joe V and Brown, Jeffrey S},
  journal={Journal of the American Medical Informatics Association},
  volume={21},
  number={4},
  pages={578--582},
  year={2014},
  publisher={Oxford Academic}
}

@misc{ohdsi20,
  title        = {Standardized Data: The {OMOP} Common Data Model ({OHDSI})},
  howpublished = {\url{https://www.ohdsi.org/data-standardization/}},
  year         = {2020}
}

@misc{pcornet25,
  title        = {{PCORnet}: The National Patient-Centered Clinical Research Network},
  howpublished = {\url{https://pcornet.org/}},
  year         = {2025}
}

@article{khan2025privacy,
  title={Privacy Enhancing Technologies for Intelligent Healthcare: Research Challenges and Opportunities},
  author={Khan, Fawad and Shah, Syed Aziz and Tahir, Shahzaib and Yasin Ghadi, Yazeed and Shah, Syed Ikram and Zahid, Adnan and Ahmad, Jawad and Hussain Abbasi, Qammer},
  journal={ACM Computing Surveys},
  year={2025},
  publisher={ACM New York, NY}
}

@inproceedings{paju2023sok,
  title={{Sok}: A systematic review of {TEE} usage for developing trusted applications},
  author={Paju, Arttu and Javed, Muhammad Owais and Nurmi, Juha and Savim{\"a}ki, Juha and McGillion, Brian and Brumley, Billy Bob},
  booktitle={Proceedings of the 18th International Conference on Availability, Reliability and Security},
  pages={1--15},
  year={2023}
}

@article{geppert2022trusted,
  title={Trusted execution environments: Applications and organizational challenges},
  author={Geppert, Tim and Deml, Stefan and Sturzenegger, David and Ebert, Nico},
  journal={Frontiers in Computer Science},
  volume={4},
  pages={930741},
  year={2022},
  publisher={Frontiers Media SA}
}

@inproceedings{truex2019hybrid,
  author    = {Truex, Stacey and Baracaldo, Nathalie and Anwar, Ali and Steinke, Thomas and Ludwig, Heiko and Zhang, Rui and Zhou, Yi},
  title     = {A Hybrid Approach to Privacy-Preserving Federated Learning},
  booktitle = {Proceedings of the 12th ACM Workshop on Artificial Intelligence and Security (AISec)},
  year      = {2019},
  pages     = {1--11},
}

@inproceedings{bonawitz2017secureagg,
  author    = {Keith Bonawitz and Vladimir Ivanov and Ben Kreuter and Antonio Marcedone and H. Brendan {McMahan} and Sarvar Patel and Daniel Ramage and Aaron Segal and Karn Seth},
  title     = {Practical Secure Aggregation for Privacy-Preserving Machine Learning},
  booktitle = {Proceedings of the 2017 ACM SIGSAC Conference on Computer and Communications Security (CCS)},
  year      = {2017},
  pages     = {1175--1191},
  doi       = {10.1145/3133956.3133982}
}

@article{bonawitz2019towards,
  title={Towards federated learning at scale: System design},
  author={Bonawitz, Keith and Eichner, Hubert and Grieskamp, Wolfgang and Huba, Dzmitry and Ingerman, Alex and Ivanov, Vladimir and Kiddon, Chloe and Kone{\v{c}}n{\`y}, Jakub and Mazzocchi, Stefano and {McMahan}, Brendan and others},
  journal={Proceedings of machine learning and systems (MLSys)},
  volume={1},
  pages={374--388},
  year={2019}
}

@inproceedings{mcmahan2017communication,
  title={Communication-efficient learning of deep networks from decentralized data},
  author={{McMahan}, Brendan and Moore, Eider and Ramage, Daniel and Hampson, Seth and y Arcas, Blaise Aguera},
  booktitle={Artificial Intelligence and Statistics (AISTATS)},
  pages={1273--1282},
  year={2017},
  organization={PMLR}
}

@article{froelicher21,
  author  = {Dimitrios Froelicher and Juan R. Troncoso-Pastoriza and Jean-Philippe Bossuat and Marcello H{\"a}hnel and Vasileios Belagiannis and Pascal Frossard and Antoine Geissb{\"u}hler and Jean-Pierre Hubaux},
  title   = {Truly Privacy-Preserving Federated Analytics for Precision Medicine with Multiparty Homomorphic Encryption},
  journal = {Nature Communications},
  year    = {2021},
  volume  = {12},
  number  = {1},
  pages   = {5910},
  doi     = {10.1038/s41467-021-25972-y}
}

@incollection{kasivis2013nodeDP,
  author    = {Shiva Prasad Kasiviswanathan and Kobbi Nissim and Sofya Raskhodnikova and Adam Smith},
  title     = {Analyzing Graphs with Node Differential Privacy},
  booktitle = {Theory of Cryptography Conference (TCC 2013)},
  publisher = {Springer},
  year      = {2013},
  pages     = {457--476},
  doi       = {10.1007/978-3-642-36594-2_26}
}

@inproceedings{hay2009graphdp,
  author    = {Michael Hay and Chao Li and Gerome Miklau and David Jensen},
  title     = {Accurate Estimation of the Degree Distribution of Private Networks},
  booktitle = {2009 IEEE International Conference on Data Mining (ICDM)},
  year      = {2009},
  pages     = {169--178},
  doi       = {10.1109/ICDM.2009.11}
}

@article{rogers2016privacy,
  title={Privacy odometers and filters: Pay-as-you-go composition},
  author={Rogers, Ryan M and Roth, Aaron and Ullman, Jonathan and Vadhan, Salil},
  journal={Advances in Neural Information Processing Systems (NeurIPS)},
  volume={29},
  year={2016}
}

@article{khalid2023privacy,
  title={Privacy-preserving artificial intelligence in healthcare: Techniques and applications},
  author={Khalid, Nazish and Qayyum, Adnan and Bilal, Muhammad and Al-Fuqaha, Ala and Qadir, Junaid},
  journal={Computers in Biology and Medicine},
  volume={158},
  pages={106848},
  year={2023},
  publisher={Elsevier}
}

@article{ficek2021differential,
  title={Differential privacy in health research: A scoping review},
  author={Ficek, Joseph and Wang, Wei and Chen, Henian and Dagne, Getachew and Daley, Ellen},
  journal={Journal of the American Medical Informatics Association},
  volume={28},
  number={10},
  pages={2269--2276},
  year={2021},
  publisher={Oxford University Press}
}

@article{uhler2013privgenomics,
  author  = {Caroline Uhler and Aleksandra Slavkovi{\'c} and Stephen E. Fienberg},
  title   = {Privacy-Preserving Data Sharing for Genome-Wide Association Studies},
  journal = {Journal of Privacy and Confidentiality},
  year    = {2013},
  volume  = {5},
  number  = {1},
  pages   = {137--166}
}

@misc{pdasp24,
  title        = {Privacy-Preserving Data Sharing in Practice ({PDaSP}), {NSF} 24-585},
  howpublished = {National Science Foundation, Program Solicitation},
  year         = {2024},
  url          = {https://www.nsf.gov/funding/opportunities/pdasp-privacy-preserving-data-sharing-practice/506327/nsf24-585}
}

@article{tramer2020adaptive,
  title={On adaptive attacks to adversarial example defenses},
  author={Tramer, Florian and Carlini, Nicholas and Brendel, Wieland and Madry, Aleksander},
  journal={Advances in neural information processing systems},
  volume={33},
  pages={1633--1645},
  year={2020}
}

@inproceedings{ginart2019making,
  author    = {Antonio Ginart and Melody Guan and Gregory Valiant and James Y Zou},
  title     = {Making {AI} Forget You: Data Deletion in Machine Learning},
  booktitle = {Advances in neural information processing systems (NeurIPS)},
  year      = {2019},
  volume    = {32},
  pages     = {3518--3531}
}

@inproceedings{neel2021descent,
  title={Descent-to-delete: Gradient-based methods for machine unlearning},
  author={Neel, Seth and Roth, Aaron and Sharifi-Malvajerdi, Saeed},
  booktitle={Algorithmic Learning Theory (ALT)},
  pages={931--962},
  year={2021},
  organization={PMLR}
}

@article{papachristou_rahimian_dpdi_medrxiv_2025,
  author  = {Marios Papachristou and M. Amin Rahimian},
  title   = {Differentially Private Distributed Inference},
  journal = {medRxiv},
  year    = {2025},
  month   = mar,
  doi     = {10.1101/2025.03.10.25323686},
  url     = {https://www.medrxiv.org/content/10.1101/2025.03.10.25323686},
  note    = {Preprint}
}

@article{rahimian2023seeding,
  title        = {Seeding with Differentially Private Network Information},
  author       = {Liu, Yuxin and Rahimian, M. Amin and Yu, Fang{-}Yi},
  journal      = {arXiv},
  eprint       = {2305.16590},
  archivePrefix= {arXiv},
  primaryClass = {cs.SI},
  year         = {2025},
  url          = {https://arxiv.org/abs/2305.16590}
}

@inproceedings{rahimian2023aamas,
  title        = {Differentially Private Network Data Collection for Influence Maximization},
  author       = {Rahimian, M. Amin and Yu, Fang{-}Yi and Hurtado, Carlos},
  booktitle    = {Proceedings of the 22nd International Conference on Autonomous Agents and Multiagent Systems (AAMAS 2023)},
  pages        = {2795--2797},
  year         = {2023},
  publisher    = {International Foundation for Autonomous Agents and Multiagent Systems},
  url          = {https://dl.acm.org/doi/10.5555/3545946.3599081}
}

@article{weiss2020artnet,
  title        = {Egocentric Sexual Networks of Men Who Have Sex with Men in the United States: Results from the {ARTnet} Study},
  author = {Weiss, Kevin M. and Goodreau, Steven M. and Morris, Martina and others},
  journal      = {Epidemics},
  volume       = {30},
  pages        = {100386},
  year         = {2020},
  doi          = {10.1016/j.epidem.2020.100386},
  url          = {https://pubmed.ncbi.nlm.nih.gov/32004795/}
}

@article{juusola2012prep,
  title        = {The Cost-Effectiveness of Pre-Exposure Prophylaxis for {HIV} Prevention in the United States in Men Who Have Sex with Men},
  author       = {Juusola, Jessie L. and Brandeau, Margaret L. and Owens, Douglas K. and Bendavid, Eran},
  journal      = {Annals of Internal Medicine},
  volume       = {156},
  number       = {8},
  pages        = {541--550},
  year         = {2012},
  doi          = {10.7326/0003-4819-156-8-201204170-00001},
  url          = {https://pmc.ncbi.nlm.nih.gov/articles/PMC3690921/}
}

@article{kessler2014prep,
  title        = {Evaluating the Impact of Prioritization of Antiretroviral Pre-Exposure Prophylaxis in New York City},
  author       = {Kessler, Jason and Myers, Jessica E. and Nucifora, Kathryn and Mensah, Nicholas and Toohey, Ciann and Khademi, Amir and Schackman, Bruce R. and Braunstein, Sarah L.},
  journal      = {AIDS},
  volume       = {28},
  number       = {18},
  pages        = {2683--2691},
  year         = {2014},
  doi          = {10.1097/QAD.0000000000000460},
  url          = {https://pmc.ncbi.nlm.nih.gov/articles/PMC4556593/}
}

@article{tian2024reliable,
  title={Reliable generation of privacy-preserving synthetic electronic health record time series via diffusion models},
  author={Tian, Muhang and Chen, Bernie and Guo, Allan and Jiang, Shiyi and Zhang, Anru R},
  journal={Journal of the American Medical Informatics Association},
  volume={31},
  number={11},
  pages={2529--2539},
  year={2024},
  publisher={Oxford University Press}
}

@article{adnan2022scirep,
  title   = {Federated learning with differential privacy for histopathology image analysis},
  author  = {Adnan, Muhammad and Kalra, Saira and Cresswell, James C. and Taylor, Gary W. and Tizhoosh, Hamid R.},
  journal = {Scientific Reports},
  year    = {2022},
  volume  = {12},
  number  = {1},
  pages   = {12508},
  doi     = {10.1038/s41598-022-09502-4}
}

@MISC{ai-bill-of-rights-2022,
  title        = {{Blueprint for an AI Bill of Rights: Making Automated Systems Work for the American People}},
  author       = {{White House Office of Science and Technology Policy}},
  month        = {October},
  year         = {2022},
  howpublished = {\url{https://bidenwhitehouse.archives.gov/ostp/ai-bill-of-rights/}}
}

@MISC{federal-cyber-rd-2023,
  title        = {Federal Cybersecurity Research and Development Strategic Plan},
  author       = {{National Science and Technology Council, Subcommittee on Networking and Information Technology Research and Development}},
  year         = {2023},
  howpublished = {\url{https://bidenwhitehouse.archives.gov/wp-content/uploads/2024/01/Federal-Cybersecurity-RD-Strategic-Plan-2023.pdf}}
}

@MISC{nprs-2025,
  title        = {National Privacy Research Strategy},
  author       = {{National Science and Technology Council, Subcommittee on Networking and Information Technology Research and Development}},
  year         = {2025},
  howpublished = {\url{https://www.nitrd.gov/pubs/National-Privacy-Research-Strategy-2025.pdf}}
}

@BOOK{dwork14,
  title     = {The Algorithmic Foundations of Differential Privacy},
  author    = {Cynthia Dwork and Aaron Roth},
  year      = {2014},
  publisher = {Now Publishers}
}

@MISC{ai-eo,
  title        = "{Executive Order on the Safe, Secure, and Trustworthy Development and Use of Artificial Intelligence}",
  howpublished = "\url{https://bidenwhitehouse.archives.gov/briefing-room/presidential-actions/2023/10/30/executive-order-on-the-safe-secure-and-trustworthy-development-and-use-of-artificial-intelligence/}",
  note         = "The White House",
  month        = "October",
  year         = "2023",
}

@techreport{national-strategy-ppdsa,
  author       = {{National Science and Technology Council, Subcommittee on Privacy Enhancing Technologies}},
  title        = {National Strategy to Advance Privacy-Preserving Data Sharing and Analytics},
  institution  = {Executive Office of the President},
  year         = {2023},
  howpublished = {\url{https://www.nitrd.gov/pubs/National-Strategy-to-Advance-Privacy-Preserving-Data-Sharing-and-Analytics.pdf}}
}

@misc{appleDP,
  author       = {{Apple Differential Privacy Team}},
  title        = {Learning with Privacy at Scale},
  year         = {2017},
  howpublished = {\url{https://machinelearning.apple.com/research/learning-with-privacy-at-scale}}
}

@misc{erlingsson2014learning,
  author       = {Ulfar Erlingsson},
  title        = {Learning Statistics with Privacy, Aided by the Flip of a Coin},
  year         = {2014},
  month        = {Oct},
  howpublished = {\url{https://security.googleblog.com/2014/10/learning-statistics-with-privacy-aided.html}}
}

@article{barbaro2006face,
  author  = {Barbaro, Michael and Zeller, Tom and Hansell, Saul},
  title   = {A Face Is Exposed for {AOL} Searcher No. 4417749},
  journal = {The New York Times},
  year    = {2006},
  month   = {Aug},
  note    = {\url{https://www.nytimes.com/2006/08/09/technology/09aol.html}}
}

@inproceedings{kumar2007anonymizing,
  author    = {Kumar, Ravi and Novak, Jasmine and Pang, Bo and Tomkins, Andrew},
  title     = {On Anonymizing Query Logs via Token-Based Hashing},
  booktitle = {Proceedings of the 16th International World Wide Web Conference (WWW)},
  year      = {2007},
  pages     = {629--638},
  publisher = {ACM}
}

@article{sweeney1997weaving,
  author  = {Sweeney, Latanya},
  title   = {Weaving Technology and Policy Together to Maintain Confidentiality},
  journal = {The Journal of Law, Medicine \& Ethics},
  year    = {1997},
  volume  = {25},
  number  = {2-3},
  pages   = {98--110},
  doi     = {10.1111/j.1748-720X.1997.tb01885.x}
}

@article{sweeney2015only,
  author  = {Sweeney, Latanya},
  title   = {Only You, Your Doctor, and Many Others May Know},
  journal = {Technology Science},
  year    = {2015},
  volume  = {2015092903},
  note    = {\url{https://techscience.org/a/2015092903/}}
}

@article{gong2022harnessing,
  title={Harnessing the known unknowns: Differential privacy and the 2020 census},
  author={Gong, Ruobin and Groshen, Erica L and Vadhan, Salil},
  journal={Harvard Data Science Review},
  number={Special Issue 2},
  year={2022},
  publisher={The MIT Press}
}

@inproceedings{hod2023designing,
  author    = {Hod, Shlomi},
  title     = {Designing the Pilot Release of {Israel}’s National Registry of Live Births: Reconciling Privacy with Accuracy and Usability},
  booktitle = {USENIX Conference on Privacy Engineering Practice and Respect (PEPR)},
  year      = {2023},
  note      = {\url{https://www.usenix.org/conference/pepr23/presentation/hod}}
}

@inproceedings{hod2025differentially,
  title={Differentially private release of {Israel}'s national registry of live births},
  author={Hod, Shlomi and Canetti, Ran},
  booktitle={2025 IEEE Symposium on Security and Privacy (SP)},
  pages={3912--3930},
  year={2025},
  organization={IEEE}
}

@article{kairouz21,
  author  = {Kairouz, Peter and {McMahan}, H. Brendan and others},
  title   = {Advances and Open Problems in Federated Learning},
  journal = {Foundations and Trends in Machine Learning},
  year    = {2021},
  volume  = {14},
  number  = {1--2},
  pages   = {1--210},
  doi     = {10.1561/2200000083},
}

@incollection{nass2009value,
  title     = {The value and importance of health information privacy},
  author    = {Nass, Sharyl J and Levit, Laura A and Gostin, Lawrence O and others},
  booktitle = {Beyond the HIPAA Privacy Rule: Enhancing Privacy, Improving Health Through Research},
  year      = {2009},
  publisher = {National Academies Press (US)}
}

@misc{NIH_DMS_2021,
  title   = {Final {NIH} Policy for Data Management and Sharing ({NOT}-{OD}-21-{013})},
  author  = {{National Institutes of Health}},
  year    = {2020},
  url     = {https://grants.nih.gov/grants/guide/notice-files/NOT-OD-21-013.html},
  note    = {Effective Jan 25, 2023},
}

@misc{NIH_DMS_Reminder_2023,
  title   = {Reminder: {NIH} Policy for Data Management and Sharing effective on January 25, 2023 ({NOT-OD-23-053})},
  author  = {{National Institutes of Health}},
  year    = {2023},
  url     = {https://grants.nih.gov/grants/guide/notice-files/NOT-OD-23-053.html}
}

@misc{FTC_HBNR_PR_2024,
  title   = {{FTC} finalizes changes to the Health Breach Notification Rule},
  author  = {{Federal Trade Commission}},
  year    = {2024},
  month   = {April},
  url     = {https://www.ftc.gov/news-events/news/press-releases/2024/04/ftc-finalizes-changes-health-breach-notification-rule}
}

@misc{FTC_HBNR_FR_2024,
  title   = {Health Breach Notification Rule — Final Rule ({Apr}. 25, 2024)},
  author  = {{Federal Trade Commission}},
  year    = {2024},
  url     = {https://www.ftc.gov/system/files/ftc_gov/pdf/hbnr_final_rule_04_25.pdf}
}

@misc{FTC_HBNR_Basics_2024,
  title   = {Health Breach Notification Rule: The Basics for Business (updated {July} 2024)},
  author  = {{Federal Trade Commission}},
  year    = {2024},
  url     = {https://www.ftc.gov/business-guidance/resources/health-breach-notification-rule-basics-business}
}

@misc{ONC_HTI1_FR_2024,
  title   = {Health Data, Technology, and Interoperability: Certification Program Updates, Algorithm Transparency, and Information Sharing ({HTI}\textendash1) — Final Rule},
  author  = {{Office of the National Coordinator for Health IT}},
  year    = {2024},
  month   = {January},
  howpublished = {Federal Register},
  url     = {https://www.federalregister.gov/documents/2024/01/09/2023-28857/health-data-technology-and-interoperability-certification-program-updates-algorithm-transparency-and}
}

@misc{HTI1_Explainers_2024,
  title   = {{HTI}\textendash1 Final Rule resources on predictive decision support and information sharing},
  author  = {{HIMSS and allied sources}},
  year    = {2024},
  url     = {https://www.himss.org/resources/hti-1-final-rule-five-key-resources-understanding-new-standards-specifications}
}

@misc{TEFCA_QHINs,
  title   = {Designated {QHIN}s},
  author  = {{RCE / The Sequoia Project}},
  year    = {2025},
  url     = {https://rce.sequoiaproject.org/designated-qhins/}
}

@misc{HIPAA_Repro_FactSheet_2025,
  title   = {{HIPAA} Privacy Rule to Support Reproductive Health Care Privacy — Final Rule Fact Sheet},
  author  = {{{US} Department of Health and Human Services, OCR}},
  year    = {2025},
  url     = {https://www.hhs.gov/hipaa/for-professionals/special-topics/reproductive-health/final-rule-fact-sheet/index.html}
}

@misc{HIPAA_Repro_Vacatur_2025,
  title   = {{Purl v. HHS} ({N.D.} {Tex.} {June} 18, 2025) — Order vacating the Reproductive Health Privacy Rule},
  author  = {{Holland \& Knight summary}},
  year    = {2025},
  url     = {https://www.hklaw.com/en/insights/publications/2025/06/hipaas-reproductive-health-rule-is-vacated-nationally}
}

@misc{Reuters_Vacatur_2025,
  title   = {US judge invalidates Biden rule protecting privacy for abortions},
  author  = {Reuters},
  year    = {2025},
  month   = {June},
  url     = {https://www.reuters.com/business/healthcare-pharmaceuticals/us-judge-invalidates-biden-rule-protecting-privacy-abortions-2025-06-18/}
}

@misc{EU_AI_Act_OJ_2024,
  title   = {Regulation (EU) 2024/1689 — Artificial Intelligence Act (Official Journal, 12 July 2024)},
  author  = {{European Union}},
  year    = {2024},
  url     = {https://eur-lex.europa.eu/eli/reg/2024/1689/oj/eng}
}

@misc{MRCT_GDPR_2024,
  title   = {Impact of {GDPR} and Privacy Laws on Clinical Research},
  author  = {{Multi-Regional Clinical Trials (MRCT) Center}},
  year    = {2024},
  url     = {https://mrctcenter.org/project/impact-of-gdpr-and-privacy-laws-on-clinical-research/}
}

@misc{CA_CPRA_SPI_1798_140,
  title  = {California Civil Code § 1798.140 - Definitions ({CCPA}/{CPRA})},
  author = {{California Legislature}},
  year   = {2018},
  url    = {https://leginfo.legislature.ca.gov/faces/codes_displaySection.xhtml?lawCode=CIV&sectionNum=1798.140},
}

@misc{WA_MHMDA,
  title  = {Washington My Health My Data Act, Chapter 19.373 {RCW}},
  author = {{Washington State Legislature}},
  year   = {2023},
  url    = {https://app.leg.wa.gov/RCW/default.aspx?cite=19.373&full=true}
}

@misc{WA_Biometric_Identifiers,
  title  = {Biometric Identifiers, Chapter 19.375 {RCW}},
  author = {{Washington State Legislature}},
  year   = {2017},
  url    = {https://app.leg.wa.gov/RCW/default.aspx?cite=19.375&full=true}
}

@misc{BIPA_SB2979_2024,
  title  = {Illinois {BIPA} Amendments ({SB 2979, Aug.\ 2, 2024}): Per-person damages and electronic consent},
  author = {{State of Illinois}},
  year   = {2024},
  url    = {https://www.gtlaw.com/en/insights/2024/8/bipa-update-illinois-limits-liability-and-clarifies-electronic-consent-for-biometric-data-collection}
}

@misc{TX_Meta_1p4B_2024,
  title  = {Texas {AG} announces \$1.4B settlement with Meta over biometric data practices},
  author = {{Office of the Texas Attorney General}},
  year   = {2024},
  month  = {July},
  url    = {https://www.texasattorneygeneral.gov/news/releases/attorney-general-ken-paxton-secures-14-billion-settlement-meta-over-its-unauthorized-capture}
}

@misc{TX_Google_1p4B_2025,
  title  = {Google to pay \$1.4B to {Texas} in biometric and data-privacy settlement},
  author = {{Associated Press}},
  year   = {2025},
  month  = {May},
  url    = {https://apnews.com/article/8097e181cc7cb8522781db8a9a897eea}
}

@misc{FTC_Biometric_Policy_2023,
  title  = {Commission Policy Statement on Biometric Information},
  author = {{Federal Trade Commission}},
  year   = {2023},
  month  = {May},
  url    = {https://www.ftc.gov/system/files/ftc_gov/pdf/p225402biometricpolicystatement.pdf}
}

@misc{APRA_Sensitive_2024,
  title  = {Understanding `sensitive covered data' under the {APRA} discussion draft},
  author = {{IAPP}},
  year   = {2024},
  month  = {June},
  url    = {https://iapp.org/news/a/understanding-sensitive-covered-data-under-the-apra-discussion-draft}
}

@misc{TPPA_2025,
  title  = {Traveler Privacy Protection Act of 2025 ({S.1691}) — bill text},
  author = {{US Congress}},
  year   = {2025},
  month  = {May},
  url    = {https://www.congress.gov/bill/119th-congress/senate-bill/1691/text}
}

@article{owsley_cross-sectional_2025,
    title = {Cross-sectional design and protocol for {Artificial} {Intelligence} {Ready} and {Equitable} {Atlas} for {Diabetes} {Insights} ({AI}-{READI})},
    volume = {15},
    issn = {2044-6055, 2044-6055},
    url = {https://bmjopen.bmj.com/lookup/doi/10.1136/bmjopen-2024-097449},
    doi = {10.1136/bmjopen-2024-097449},
    language = {en},
    number = {2},
    urldate = {2025-11-02},
    journal = {BMJ Open},
    author = {Owsley, Cynthia and Matthies, Dawn S and McGwin, Gerald and Edberg, Jeffrey C and Baxter, Sally L and Zangwill, Linda M and Owen, Julia P and Lee, Cecilia S},
    month = feb,
    year = {2025},
    pages = {e097449},
}

@misc{hallaj_open_2025,
    title = {Open {Data} {Sharing} in {Clinical} {Research} and {Participants} {Privacy}: {Challenges} and {Opportunities} in the {Era} of {Artificial} {Intelligence}},
    shorttitle = {Open {Data} {Sharing} in {Clinical} {Research} and {Participants} {Privacy}},
    url = {http://arxiv.org/abs/2508.01140},
    doi = {10.48550/arXiv.2508.01140},
    urldate = {2025-11-02},
    publisher = {arXiv},
    author = {Hallaj, Shahin and Heinke, Anna and Kalaw, Fritz Gerald P. and Gim, Nayoon and Blazes, Marian and Owen, Julia and Dysinger, Eamon and Benton, Erik S. and Cordier, Benjamin A. and Evans, Nicholas G. and Li-Pook-Than, Jennifer and Snyder, Michael P. and Nebeker, Camille and Zangwill, Linda M. and Baxter, Sally L. and McWeeney, Shannon and Lee, Cecilia S. and Lee, Aaron Y. and Patel, Bhavesh},
    month = aug,
    year = {2025},
    note = {arXiv:2508.01140 [q-bio]},
}

@article{homer2008resolving,
  title={Resolving individuals contributing trace amounts of {DNA} to highly complex mixtures using high-density {SNP} genotyping microarrays},
  author={Homer, Nils and Szelinger, Szabolcs and Redman, Margot and Duggan, David and Tembe, Waibhav and Muehling, Jill and Pearson, John V and Stephan, Dietrich A and Nelson, Stanley F and Craig, David W},
  journal={PLoS genetics},
  volume={4},
  number={8},
  pages={e1000167},
  year={2008},
  publisher={Public Library of Science San Francisco, USA}
}

@article{Gymrek2013Science,
  author  = {Gymrek, Melissa and McGuire, Amy L. and Golan, David and Halperin, Eran and Erlich, Yaniv},
  title   = {Identifying Personal Genomes by Surname Inference},
  journal = {Science},
  year    = {2013},
  volume  = {339},
  number  = {6117},
  pages   = {321--324},
  doi     = {10.1126/science.1229566}
}

@techreport{pulivarti2024genomic,
  title={Genomic Data Cybersecurity and Privacy Frameworks Community Profile},
  author={Pulivarti, Ronald and Wagner, Justin and Zook, Justin and Craft, R and Kreider, Brett and Miller, Jeremy and O'Neil, Patrick and Sames, Christina and Snyder, Julie and Stea, Bob and others},
  year={2024},
  institution={National Institute of Standards and Technology}
}

@inproceedings{Qiao2021ICML,
  author    = {Qiao, Guanghao and Su, Weijie J. and Zhang, Linjun},
  title     = {One-shot Differentially Private Top-$k$ Selection},
  booktitle = {Proceedings of the 38th International Conference on Machine Learning},
  year      = {2021},
  pages     = {8672--8681},
  publisher = {PMLR}
}

@article{Almadhoun2020Bioinformatics,
  author  = {Almadhoun, Nour and Ayday, Erman and Ulusoy, {\"O}zg{\"u}r},
  title   = {Differential privacy under dependent tuples: The case of genomic privacy},
  journal = {Bioinformatics},
  year    = {2020},
  volume  = {36},
  number  = {6},
  pages   = {1696--1703},
  doi     = {10.1093/bioinformatics/btz791}
}

@inproceedings{Yilmaz2022CODASPY,
  author    = {Yilmaz, Erman and Ji, Tianhao and Ayday, Erman and Li, Pan},
  title     = {Genomic Data Sharing under Dependent Local Differential Privacy},
  booktitle = {Proceedings of the 12th ACM Conference on Data and Application Security and Privacy (CODASPY)},
  year      = {2022},
  pages     = {77--88},
  publisher = {ACM}
}

@article{cannon2024digital,
  title   = {Digital twin mathematical models suggest individualized hemorrhagic shock resuscitation strategies},
  author  = {Cannon, Jeremy W and Gruen, Danielle S and Zamora, Ruben and Brostoff, Noah and Hurst, Kelly and Harn, John H and El-Dehaibi, Fayten and Geng, Zhi and Namas, Rami and Sperry, Jason L and Holcomb, John B and Cotton, Bryan A and Nam, Jason J and Underwood, Samantha and Schreiber, Martin A and Chung, Kevin K and Batchinsky, Andriy I and Cancio, Leopoldo C and Benjamin, Andrew J and Fox, Erin E and Chang, Steven C and Cap, Andrew P and Vodovotz, Yoram},
  journal = {Communications Medicine},
  year    = {2024},
  volume  = {4},
  number  = {1},
  pages   = {113},
  doi     = {10.1038/s43856-024-00535-6}
}

@article{chadebecq2023artificial,
  title={Artificial intelligence and automation in endoscopy and surgery},
  author={Chadebecq, Fran{\c{c}}ois and Lovat, Laurence B and Stoyanov, Danail},
  journal={Nature Reviews Gastroenterology \& Hepatology},
  volume={20},
  number={3},
  pages={171--182},
  year={2023},
  publisher={Nature Publishing Group UK London}
}

@article{chanda2024dermatologist,
  title   = {Dermatologist-like explainable {AI} enhances trust and confidence in diagnosing melanoma},
  author  = {Chanda, Tirtha and Hauser, Katja and Hobelsberger, Sarah and Bucher, Tabea-Clara and Nogueira Garcia, Carina and Wies, Christoph and Kittler, Harald and Tschandl, Philipp and Navarrete-Dechent, Cristian and Podlipnik, Sebastian and Chousakos, Emmanouil and Crnaric, Iva and Majstorovic, Jovana and Alhajwan, Linda and Foreman, Tanya and Peternel, Sandra and Sarap, Sergei and {\"O}zdemir, {\.I}rem and Barnhill, Raymond L and Llamas-Velasco, Mar and Poch, Gabriela and Korsing, S{\"o}ren and Sondermann, Wiebke and Gellrich, Frank Friedrich and Heppt, Markus V and Erdmann, Michael and Haferkamp, Sebastian and Drexler, Konstantin and Goebeler, Matthias and Schilling, Bastian and Utikal, Jochen S and Ghoreschi, Kamran and Fr{\"o}hling, Stefan and Krieghoff-Henning, Eva and Reader Study Consortium and Brinker, Titus J},
  journal = {Nature Communications},
  year    = {2024},
  volume  = {15},
  number  = {1},
  pages   = {524},
  doi     = {10.1038/s41467-023-43095-4}
}

@article{chen2023digital,
  title={Digital health for aging populations},
  author={Chen, Chuanrui and Ding, Shichao and Wang, Joseph},
  journal={Nature Medicine},
  volume={29},
  number={7},
  pages={1623--1630},
  year={2023},
  publisher={Nature Publishing Group US New York}
}

@article{esteva2017dermatologist,
  title={Dermatologist-level classification of skin cancer with deep neural networks},
  author={Esteva, Andre and Kuprel, Brett and Novoa, Roberto A and Ko, Justin and Swetter, Susan M and Blau, Helen M and Thrun, Sebastian},
  journal={Nature},
  volume={542},
  number={7639},
  pages={115--118},
  year={2017}
}

@article{ferretti2020quantifying,
  title={Quantifying {SARS-CoV-2} transmission suggests epidemic control with digital contact tracing},
  author={Ferretti, Luca and Wymant, Chris and Kendall, Michelle and Zhao, Lele and Nurtay, Anel and Abeler-D{\"o}rner, Lucie and Parker, Michael and Bonsall, David and Fraser, Christophe},
  journal={Science},
  volume={368},
  number={6491},
  pages={eabb6936},
  year={2020}
}

@article{ferretti2024digital,
  title={Digital measurement of {SARS-CoV-2} transmission risk from 7 million contacts},
  author={Ferretti, Luca and Wymant, Chris and Petrie, James and Tsallis, Daphne and Kendall, Michelle and Ledda, Alice and Di Lauro, Francesco and Fowler, Adam and Di Francia, Andrea and Panovska-Griffiths, Jasmina and others},
  journal={Nature},
  volume={626},
  number={7997},
  pages={145--150},
  year={2024},
  publisher={Nature Publishing Group UK London}
}

@article{gargeya2017automated,
  title={Automated identification of diabetic retinopathy using deep learning},
  author={Gargeya, Rishab and Leng, Theodore},
  journal={Ophthalmology},
  volume={124},
  number={7},
  pages={962--969},
  year={2017}
}

@article{hannun2019cardiologist,
  title={Cardiologist-level arrhythmia detection and classification in ambulatory electrocardiograms using a deep neural network},
  author={Hannun, Awni Y and Rajpurkar, Pranav and Haghpanahi, Masoumeh and Tison, Geoffrey H and Bourn, Codie and Turakhia, M P and Ng, Andrew Y},
  journal={Nature Medicine},
  volume={25},
  number={1},
  pages={65--69},
  year={2019}
}

@article{hodson2016precision,
  title={Precision medicine},
  author={Hodson, R},
  journal={Nature},
  volume={537},
  number={7619},
  pages={S49--S49},
  year={2016}
}

@article{kuan2019adopting,
  title={Adopting {AI} in health care will be slow and difficult},
  author={Kuan, Roger},
  journal={Harvard Business Review},
  volume={10},
  pages={1},
  year={2019}
}

@article{li2023systematic,
  title={Systematic review and meta-analysis of {AI}-based conversational agents for promoting mental health and well-being},
  author={Li, Han and Zhang, Renwen and Lee, Yi-Chieh and Kraut, Robert E and Mohr, David C},
  journal={NPJ Digital Medicine},
  volume={6},
  number={1},
  pages={236},
  year={2023},
  publisher={Nature Publishing Group UK London}
}

@article{miller2022privacy,
  title={Privacy of digital health information},
  author={Miller, Amalia R},
  journal={Economics of Privacy},
  year={2022},
  pages = {1077--1093},
  publisher={University of Chicago Press}
}

@inproceedings{nissenbaum2004privacy,
  title={Privacy as contextual integrity},
  author={Nissenbaum, Helen},
  booktitle={Washington Law Review},
  volume={79},
  pages={119},
  year={2004}
}

@article{prahalad2024equitable,
  title={Equitable implementation of a precision digital health program for glucose management in individuals with newly diagnosed type 1 diabetes},
  author={Prahalad, Priya and Scheinker, David and Desai, Manisha and Ding, Victoria Y and Bishop, Franziska K and Lee, Ming Yeh and Ferstad, Johannes and Zaharieva, Dessi P and Addala, Ananta and Johari, Ramesh and others},
  journal={Nature Medicine},
  pages={1--9},
  year={2024},
  publisher={Nature Publishing Group US New York}
}

@article{rajpurkar2022ai,
  title={{AI} in health and medicine},
  author={Rajpurkar, Pranav and Chen, Emma and Banerjee, Oishi and Topol, Eric J},
  journal={Nature Medicine},
  volume={28},
  number={1},
  pages={31--38},
  year={2022}
}

@article{shandhi2024assessment,
  title={Assessment of ownership of smart devices and the acceptability of digital health data sharing},
  author={Shandhi, Md Mobashir Hasan and Singh, Karnika and Janson, Natasha and Ashar, Perisa and Singh, Geetika and Lu, Baiying and Hillygus, D Sunshine and Maddocks, Jennifer M and Dunn, Jessilyn P},
  journal={NPJ Digital Medicine},
  volume={7},
  number={1},
  pages={44},
  year={2024},
  publisher={Nature Publishing Group UK London}
}

@article{wang2022privacy,
  title={Privacy protection in using artificial intelligence for healthcare: Chinese regulation in comparative perspective},
  author={Wang, Chao and Zhang, Jieyu and Lassi, Nicholas and Zhang, Xiaohan},
  journal={Healthcare},
  volume={10},
  number={10},
  pages={1878},
  year={2022},
  organization={MDPI}
}

@article{yang2024limits,
  title={The limits of fair medical imaging {AI} in real-world generalization},
  author={Yang, Yuzhe and Zhang, Haoran and Gichoya, Judy W and Katabi, Dina and Ghassemi, Marzyeh},
  journal={Nature Medicine},
  pages={1--11},
  year={2024},
  publisher={Nature Publishing Group US New York}
}

@article{rahimian2025democratic,
  author   = {Rahimian, M. Amin and Colaresi, Michael P.},
  title    = {Democratic Resilience and Sociotechnical Shocks},
  journal  = {Computational and Mathematical Organization Theory},
  year     = {2025},
  volume   = {31},
  pages    = {236--257},
  doi      = {10.1007/s10588-025-09394-5},
  url      = {https://doi.org/10.1007/s10588-025-09394-5},
  publisher= {Springer}
}

@article{henderson2011eradication,
  author  = {Henderson, Donald A.},
  title   = {The eradication of smallpox—an overview of the past, present, and future},
  journal = {Vaccine},
  year    = {2011},
  volume  = {29},
  number  = {Suppl 4},
  pages   = {D7--D9},
  doi     = {10.1016/j.vaccine.2011.06.080},
  url     = {https://doi.org/10.1016/j.vaccine.2011.06.080}
}

@article{reid2024programme,
  title={Programme Science in PEPFAR: a pathway to a sustainable HIV response},
  author={Reid, Michael JA and Bunnell, Rebecca and Dokubo, Emily Kainne and Nkengasong, John},
  journal={Journal of the International AIDS Society},
  volume={27},
  number={Suppl 2},
  pages={e26244},
  year={2024}
}

@article{cinelli2020covid,
  author  = {Cinelli, Matteo and Quattrociocchi, Walter and Galeazzi, Alessandro and Valensise, Carlo M. and Brugnoli, Emanuele and Schmidt, Ana Lucia and Zola, Paola and Zollo, Fabiana and Scala, Antonio},
  title   = {The {COVID}-19 social media infodemic},
  journal = {Scientific Reports},
  year    = {2020},
  volume  = {10},
  pages   = {16598},
  doi     = {10.1038/s41598-020-73510-5},
  url     = {https://doi.org/10.1038/s41598-020-73510-5}
}

@article{loomba2021measuring,
  author  = {Loomba, Sahil and de Figueiredo, Alexandre and Piatek, Simon J. and de Graaf, Kristen and Larson, Heidi J.},
  title   = {Measuring the impact of {COVID}-19 vaccine misinformation on vaccination intent in the UK and USA},
  journal = {Nature Human Behaviour},
  year    = {2021},
  volume  = {5},
  pages   = {337--348},
  doi     = {10.1038/s41562-021-01056-1},
  url     = {https://doi.org/10.1038/s41562-021-01056-1}
}

@article{skafle2022misinformation,
  author  = {Skafle, Ingjerd and Nordahl-Hansen, Anders and Quintana, Daniel S. and Gabarron, Elia},
  title   = {Misinformation About {COVID}-19 Vaccines on Social Media: Rapid Review},
  journal = {Journal of Medical Internet Research},
  year    = {2022},
  volume  = {24},
  number  = {8},
  pages   = {e37367},
  doi     = {10.2196/37367},
  url     = {https://doi.org/10.2196/37367}
}

@article{nahum2021erosion,
  author  = {Nahum, Ari and Drekonja, Dimitri M. and Alpern, Jonathan D.},
  title   = {The Erosion of Public Trust and {SARS-CoV-2} Vaccines—More Action Is Needed},
  journal = {Open Forum Infectious Diseases},
  year    = {2021},
  volume  = {8},
  number  = {2},
  pages   = {ofaa657},
  doi     = {10.1093/ofid/ofaa657},
  url     = {https://doi.org/10.1093/ofid/ofaa657}
}

@article{carpiano2023confronting,
  author  = {Carpiano, Richard M. and Callaghan, Timothy and Mello, Michelle M. and Maldonado, Yvonne A. and DiResta, Ren{\'e}e and Brewer, Noel T. and Clinton, Chelsea and Galvani, Alison P. and Omer, Saad B. and Schwartz, Jason L. and Sharfstein, Joshua M. and Hotez, Peter J.},
  title   = {Confronting the evolution and expansion of anti-vaccine activism in the USA in the {COVID}-19 era},
  journal = {The Lancet},
  year    = {2023},
  volume  = {401},
  pages   = {967--970},
  doi     = {10.1016/S0140-6736(23)00136-8},
  url     = {https://doi.org/10.1016/S0140-6736(23)00136-8}
}

@article{papachristou2025differentially,
  title={Differentially private distributed estimation and learning},
  author={Papachristou, Marios and Rahimian, M Amin},
  journal={IISE Transactions},
  volume={57},
  number={7},
  pages={756--772},
  year={2025},
  publisher={Taylor \& Francis}
}

@article{dwork2011firm,
  title={A firm foundation for private data analysis},
  author={Dwork, Cynthia},
  journal={Communications of the ACM},
  volume={54},
  number={1},
  pages={86--95},
  year={2011},
  publisher={ACM New York, NY, USA}
}

@inproceedings{benthall2024integrating,
  title={Integrating differential privacy and contextual integrity},
  author={Benthall, Sebastian and Cummings, Rachel},
  booktitle={Proceedings of the 2024 Symposium on Computer Science and Law},
  pages={9--15},
  year={2024}
}

@article{collis2022global,
  title={Global survey on COVID-19 beliefs, behaviours and norms},
  author={Collis, Avinash and Garimella, Kiran and Moehring, Alex and Rahimian, M Amin and Babalola, Stella and Gobat, Nina H and Shattuck, Dominick and Stolow, Jeni and Aral, Sinan and Eckles, Dean},
  journal={Nature Human Behaviour},
  volume={6},
  number={9},
  pages={1310--1317},
  year={2022},
  publisher={Nature Publishing Group UK London}
}

@article{holtz2020interdependence,
  title={Interdependence and the cost of uncoordinated responses to COVID-19},
  author={Holtz, David and Zhao, Michael and Benzell, Seth G and Cao, Cathy Y and Rahimian, Mohammad Amin and Yang, Jeremy and Allen, Jennifer and Collis, Avinash and Moehring, Alex and Sowrirajan, Tara and others},
  journal={Proceedings of the National Academy of Sciences},
  volume={117},
  number={33},
  pages={19837--19843},
  year={2020},
  publisher={National Academy of Sciences}
}

\end{document}